%% file: main.tex
\documentclass[conference]{IEEEtran}
\IEEEoverridecommandlockouts
\usepackage{cite}
\usepackage{amsmath,amssymb,amsfonts}
\usepackage{graphicx}
\usepackage{textcomp}
\usepackage{xcolor}
\usepackage[nolist]{acronym}
\def\BibTeX{{\rm B\kern-.05em{\sc i\kern-.025em b}\kern-.08em
    T\kern-.1667em\lower.7ex\hbox{E}\kern-.125emX}}
\usepackage{algorithm}
\usepackage{algorithmicx}
\usepackage{algpseudocode}
\usepackage{booktabs}
\usepackage{fontawesome}
\usepackage{multirow}
\usepackage{array}
\usepackage{tabularx}
\usepackage{subcaption}
\usepackage{stfloats}

\usepackage{newtxtext,newtxmath}

\begin{document}
\input{acronym}
\title{NeuroScaler: Towards Energy-Optimal Autoscaling for Container-Based Services}

\author{\IEEEauthorblockN{Alisson O. Chaves\textsuperscript{1}\textsuperscript{2}, Rodrigo Moreira\textsuperscript{3}, Larissa F. {Rodrigues Moreira}\textsuperscript{1}\textsuperscript{3},\\ João Correia\textsuperscript{4}, David Santos\textsuperscript{2}, Rui Silva\textsuperscript{2}, Tiago Barros\textsuperscript{2}, Daniel Corujo\textsuperscript{2}, Miguel Rocha\textsuperscript{4}\textsuperscript{5},\\ Flávio de Oliveira Silva\textsuperscript{5}}
\IEEEauthorblockA{
\textsuperscript{1}Federal University of Uberlândia (UFU), Uberlândia-MG, Brazil\\
\textsuperscript{2}Instituto de Telecomunicações (ITAV), Aveiro, Portugal\\
\textsuperscript{3}Federal University of Viçosa (UFV), Rio Paranaíba-MG, Brazil\\
\textsuperscript{4}OmniumAI, Braga, Portugal\\
\textsuperscript{5}University of Minho (UMinho), Braga, Portugal\\
Emails: \{rodrigo, larissa.f.rodrigues\}@ufv.br, jcorreia@omniumai.com,\\\{aoc.alisson, davidsantos, ruimiguelsilva, tiagobarros, dcorujo\}@av.it.pt,  \{mrocha, flavio\}@di.uminho.pt}}

\maketitle

\begin{abstract}

Future networks must meet stringent requirements while operating within tight energy and carbon constraints. Current autoscaling mechanisms remain workload-centric and infrastructure-siloed, and are largely unaware of their environmental impact. We present NeuroScaler, an AI-native, energy-efficient, and carbon-aware orchestrator for green cloud and edge networks. NeuroScaler aggregates multi-tier telemetry, from Power Distribution Units (PDUs) through bare-metal servers to virtualized infrastructure with containers managed by Kubernetes, using distinct energy and computing metrics at each tier. It supports several machine learning pipelines that link load, performance, and power. Within this unified observability layer, a model-predictive control policy optimizes energy use while meeting service-level objectives. In a real testbed with production-grade servers supporting real services, NeuroScaler reduces energy consumption by 34.68\% compared to the Horizontal Pod Autoscaler (HPA) while maintaining target latency.

\end{abstract}


\section{Introduction}\label{sec:introduction}


Sustainability~\cite{yrjola_sustainability_2020} and \ac{AI}~\cite{merluzzi_hexa-x_2023} are central to current and future networks, with \ac{AI} enabling intelligent energy management for more sustainable operations. As infrastructures move toward \ac{6G}, the growing scale of data processing and service diversity enlarges the \ac{ICT} energy footprint, making energy efficiency and carbon reduction core goals for the next decade~\cite{Martins2023}.


\ac{AI} and \ac{ML} are key enablers of sustainable mobile network management~\cite{Pivoto2023}. By combining predictive modeling, orchestration, and automation, \ac{AI}-driven frameworks adapt compute/network resources to workload changes to balance performance and energy efficiency. \ac{MLOps} advances enable continuous training, deployment, and monitoring for closed-loop control~\cite{moreira2023}.


Despite progress in carbon and energy-aware orchestration, the state of the art is still fragmented in scope and integration~\cite{Donatti2025}. Caspian and CASPER target carbon intensity and latency but stay at the cloud/application layer without telecom or multi-domain integration~\cite{caspian2024,souza_casper_2024}; CarbonScaler and Opportunistic Energy-Aware Scheduling add energy metrics and ML-based power prediction yet remain limited to data-center or single-cluster contexts~\cite{carbonscaler2024,raith2024}.



We introduce a usage-aware platform for sustainable networking that uses a modular \ac{MLOps} architecture to design, train, deploy, and monitor predictive pipelines for energy/performance optimization. By integrating heterogeneous telemetry from \acp{PDU}, hosts, \acp{VM}, and containers, it provides multi-tier visibility. It applies \ac{ML} forecasting for predictive scaling and service migration across data-center and \ac{5G}/\ac{6G} domains. Validated on a real telco-graded testbed, an \ac{AI}-native \ac{MPC} autoscaler cuts energy by 34.68\% without \acp{SLO} violations, making energy a first-class control objective across the cloud–edge–telco continuum.


The main contributions of this work are: (i) an architecture for AI-native green orchestration across the cloud–edge–\ac{5G} continuum leveraging heterogeneous power and performance telemetry; (ii) an MPC-based autoscaling framework that internalizes energy and carbon objectives without sacrificing \acp{SLO}; and (iii) experimental evidence that sustainability-aware, telco-grade automation enabled a greener digital infrastructure.


The paper is organized as follows: Section~\ref{sec:neuroscaler_sota} surveys energy-efficiency work; Section~\ref{sec:neuronet} overviews NEURONET; Section~\ref{sec:neuroscaler} details NeuroScaler; Section~\ref{sec:uc1} presents the use case; Section~\ref{sec:experiments-description} describes the experimental setup; Section~\ref{sec:results_and_discussion} reports results; Section~\ref{sec:concluding_remarks} concludes with future work.

\begin{table*}[htbp]
\scriptsize
\setlength{\tabcolsep}{3pt}
\caption{Compact qualitative comparison with the literature.}
\label{tab:neuroscaler_comparison}
\centering
\begin{tabular}{p{3.0cm} p{2.6cm} p{3.1cm} p{2.5cm} p{4.0cm}}
\toprule
\textbf{Approach} &
\textbf{Type \& Scope} &
\textbf{Orchestration Logic} &
\textbf{Use of Energy / Carbon} &
\textbf{Telco \& Multi-Domain Integration} \\
\midrule
Caspian~\cite{caspian2024} &
Carbon-aware scheduler for multi-cluster \ac{K8s}. &
Optimization-based placement using carbon intensity and capacity. &
Central design objective. &
Geo-distributed clouds only; no 5G core/RAN/transport scope. \\
\midrule
CASPER~\cite{souza_casper_2024} &
Carbon-aware provisioning for web services. &
Multi-objective optimization for emissions vs. latency \acp{SLO}. &
Central; explicit carbon minimization. &
Geo-distributed cloud and app-layer focus; no \acp{VNF}/slices integration. \\
\midrule
CarbonScaler~\cite{carbonscaler2024} &
Autoscaler for elastic batch workloads. &
Carbon-tuned scaling of batch jobs. &
Central; uses temporal carbon signals. &
Single-domain cloud; limited to flexible batch scenarios. \\
\midrule
Opportunistic Energy-Aware Scheduling~\cite{raith2024} &
Energy-aware container scheduling in DC/\ac{K8s}. &
\ac{GNN}-based power prediction plus energy-efficient placement. &
Uses power as key decision input. &
Confined to data center/\ac{K8s}; no telco or multi-domain orchestration. \\
\midrule
Spatio-Temporal Shifting~\cite{asadov2025} &
Edge-cloud carbon-aware scheduling concepts. &
Conceptual and algorithmic spatio-temporal policies. &
Carbon-centric design. &
Edge-cloud continuum; telco aspects abstracted; no concrete orchestrator. \\
\midrule
Green Cloud Continuum~\cite{patel2024} &
Analytical model for cloud-edge energy-awareness. &
Defines metrics and requirements; no runtime controller. &
Energy integral to models. &
Cloud-edge continuum; partial network view; lacks operational integration. \\
\midrule
\textbf{NEURONET NeuroScaler} &
\ac{AI}-native orchestrator across DC, edge, \ac{5G} core/\ac{RAN}, transport. &
Closed-loop, telemetry-driven placement, scaling, and migration with programmable intents. &
\textbf{Central}: multi-source energy, carbon, and \ac{QoS} jointly optimized. &
\textbf{Explicit}: unifies IT, edge, and telco domains under a single end-to-end control plane. \\
\bottomrule
\end{tabular}
\end{table*}

\section{Related Work}\label{sec:neuroscaler_sota}


This subsection surveys \ac{AI}-native, sustainability-aware orchestration in cloud/edge/telco—systems that automate deployment, placement, scaling, and migration with energy/carbon data—emphasizing works that go beyond monitoring to implement concrete control or scheduling.



Caspian is a carbon-aware scheduler for multi-cluster \ac{K8s}, placing workloads by spatio-temporal carbon intensity and resource availability~\cite{caspian2024}. It cuts emissions in geo-distributed clouds but lacks integration with \ac{5G} core, \ac{RAN}, or transport domains.


CASPER delivers carbon-aware scheduling/provisioning for geo-distributed web services via multi-objective optimization that minimizes emissions while meeting latency \acp{SLO}~\cite{souza_casper_2024}. It operates at the application/cloud-region level, without explicit coupling to telecom functions or end-to-end service chains.


CarbonScaler adjusts batch-workload resources in response to dynamic carbon-intensity signals~\cite{carbonscaler2024}. It achieves significant emission cuts in cloud batch processing but assumes flexible completion times and does not target continuous services or multi-domain orchestration.


Opportunistic Energy-Aware Scheduling~\cite{raith2024} proposes an energy-aware scheduler for container platforms that uses graph neural networks to predict host power and steer container placement. It demonstrates intra-cluster \ac{ML}-based energy gains but is limited to data-center workloads and single-domain \ac{K8s}.


Carbon-Aware Spatio-Temporal Workload Shifting~\cite{asadov2025} surveys and systematizes spatial and temporal workload shifting strategies for edge-cloud systems, proposing a carbon-aware scheduling perspective for distributed infrastructures~\cite{asadov2025}. The contribution frames the design space but does not provide a fully integrated runtime orchestrator.

Modeling the Green Cloud Continuum ~\cite{patel2024} analyzes cloud-edge continuum architectures and highlights the lack of energy awareness, proposing modeling guidelines and metrics for greener deployments~\cite{patel2024}. The work is architectural and analytical, without implementing concrete control loops for operational orchestration.


Table~\ref{tab:neuroscaler_comparison} provides a compact qualitative comparison with the literature. \emph{Type \& Scope} classifies each approach (framework/scheduler/model; cloud/edge/telco); \emph{Orchestration Logic} summarizes decision methods and lifecycle actions (deployment, placement, scaling, migration; intents/optimization/\ac{ML}); \emph{Use of Energy/Carbon Signals} notes whether these are primary, secondary, or absent; \emph{Telco \& Multi-Domain Integration} indicates coverage beyond cloud/DC to edge, \ac{5G} core, \ac{RAN}, and transport.


The related work advances intent-based automation, carbon-aware scheduling, and energy-efficient management but remains fragmented. Caspian~\cite{caspian2024}, CASPER~\cite{souza_casper_2024}, CarbonScaler~\cite{carbonscaler2024}, and Raith et al.~\cite{raith2024} center carbon/energy yet target geo-distributed web services, batch workloads, or single-domain \ac{K8s}/data-center clusters—omitting unified control across \ac{5G} core, \ac{RAN}, and transport.


Conceptual work on spatio-temporal shifting~\cite{asadov2025} and green cloud–edge continua~\cite{patel2024} clarifies sustainability goals. Still, it lacks an operational, \ac{AI}-native orchestrator that jointly reasons over data-center, edge, and network resources with fine-grained, multi-source telemetry. This exposes a gap for an integrated, multi-domain, sustainability-centric layer combining intent-driven automation, rich observability, and multi-objective (\ac{QoS}, \ac{SLA}, energy, carbon) decision-making across the cloud–edge–\ac{5G}/\ac{6G} continuum.



\section{NEURONET}\label{sec:neuronet}

\begin{figure}[H]
\centering
  \includegraphics[width=1\columnwidth]{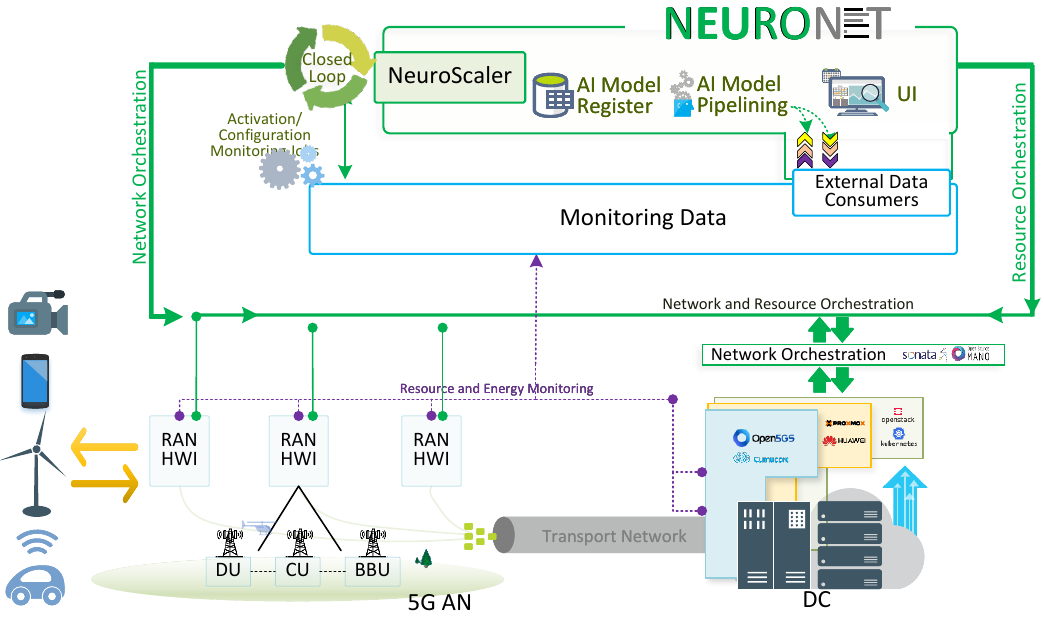}
  \caption{NEURONET overall view.}
  \label{fig:neuronet-high-view}
\end{figure}

NEURONET is an intelligent platform that contributes to sustainable and green networking by employing various \ac{AI} techniques to integrate data from the edge, the \ac{RAN}, and the data center, supporting the network core and various services. In Fig.~\ref{fig:neuronet-high-view}, we bring the overall NEURONET architecture and interactions. NEURONET handles large volumes of monitoring data, both in terms of energy and computational resources, and makes online policy decisions that minimize energy consumption while maintaining \ac{SLO}. To achieve this, NEURONET was built using \ac{MLOps} principles.

To support its goals, NEURONET introduces a software layer that integrates and enriches monitoring parameters, with a strong focus on energy-related metrics. By incorporating time series analysis with multi-step forecasting and other \ac{ML} approaches, NEURONET enables the early detection of energy consumption trends. NEURONET also includes a mechanism to feed these insights back into the network and the data center via \acp{API}, enabling proactive resource management and intelligent automation to optimize energy efficiency across the data center and network infrastructure. This work focuses on the NeuroScaler, NEURONET's orchestration entity.

\section{NeuroScaler}\label{sec:neuroscaler}

NeuroScaler orchestrates services using \ac{ML} models deployed on NEURONET. It interacts with the infrastructure to scale services up and down, start and stop them, or change their placement as needed. As the entity responsible for closed-loop integration with the infrastructure and network, it can interact with network orchestrators, \acp{VIM}, and switch devices on/off according to the infrastructure capabilities. In this work, we focus on the interaction with a Kubernetes-based infrastructure. NeuroScaler can support multiple \ac{ML} models and implement various orchestration policies based on different \ac{ML} strategies. 


\begin{figure}[H]
\centering
  \includegraphics[width=1\columnwidth]{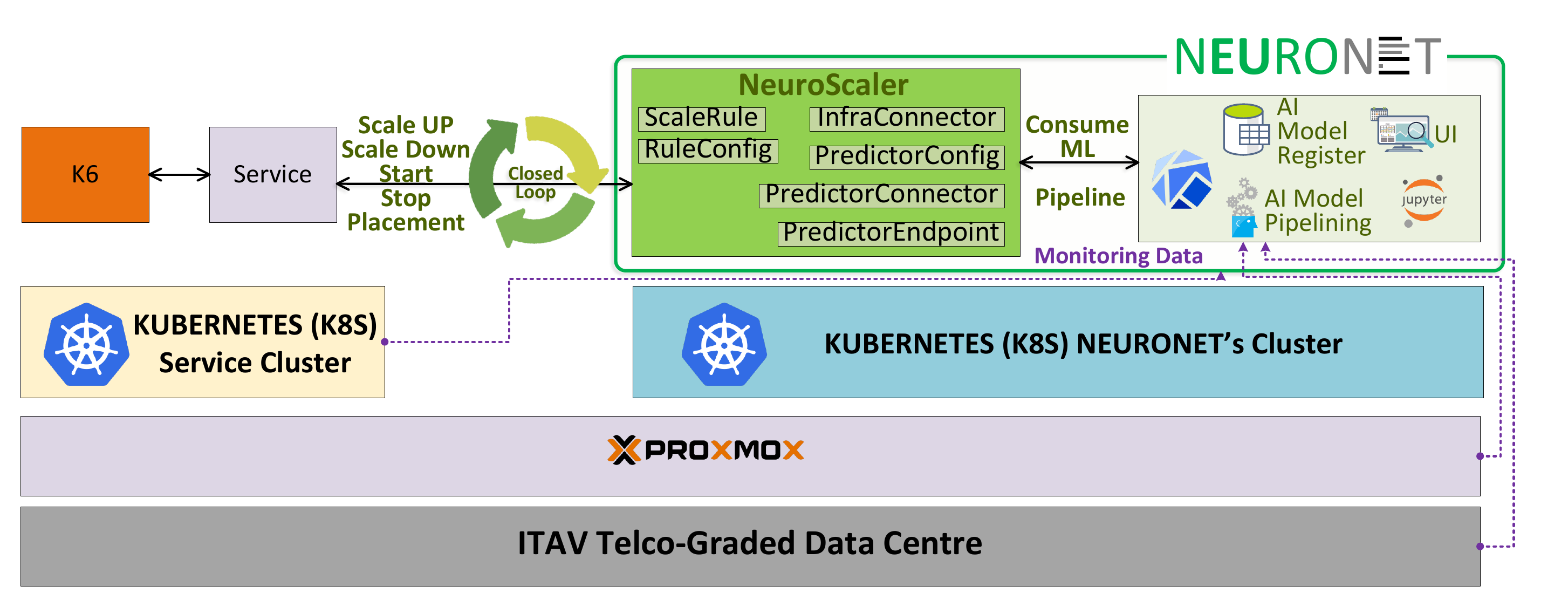}
  \caption{NeuroScaler Internal components and integrations.}
  \label{fig:neuronet-low-view}
\end{figure}


In this sense, NeuroScaler can consume various \ac{ML} pipelines, thereby supporting multiple types of predictions, such as energy consumption and service latency. 

Figure~\ref{fig:neuronet-low-view} presents the main components of NeuroScaler. Due to space constraints, we will describe only their names and roles.


\begin{itemize}
\item \textbf{ScaleRule}: Defines the autoscaling policy for predictor workloads. It specifies the predictor ID/namespace, minimum–maximum replicas, step size, and upper/lower trigger thresholds, encapsulating the logic and configuration that govern scaling behavior in \ac{K8s}.


\item \textbf{RuleConfig}: Configures and manages NeuroScaler’s scaling rules. It loads, validates, persists, and updates \texttt{ScaleRule} definitions so they’re available at runtime, while leaving decision logic to other components.


\item \textbf{PredictorEndpoint}: Represents a live predictor service endpoint (e.g., REST) that exposes runtime metrics (latency, \ac{CPU} usage, request rate). It tracks the endpoint ID, the most recent metric value, and the current status, providing feedback to inform scaling decisions.


\item \textbf{PredictorConfig}: Centralizes configuration for predictor service such as endpoint URLs, polling intervals, authentication, and supported metrics, enabling NeuroScaler to discover, connect, and authenticate to predictors for reliable, secure metric collection.


\item \textbf{InfraConnector}: Abstracts the underlying infrastructure and exposes the control plane used by NeuroScaler to read state and enact scaling. For \ac{K8s}, it stores cluster/app metadata and provides methods such as \texttt{get\_replicas()} (querying running pods) and \texttt{scale()} (adjusting replica counts), bridging high-level logic to infrastructure operations.


\item \textbf{PredictorConnector}: Implements the communication layer between NeuroScaler and deployed predictors, collecting runtime metrics or forecasts that drive scaling. It abstracts data acquisition from specific predictor implementations, providing a clean, modular interface to the scaling control system.

\end{itemize}

\subsection{Model Predictive Control (MPC) Autoscaler} \label{sec:mpc-autoscaler}

NEURONET's NeuroScaler supports multiple autoscalers. In this work, we describe the \ac{MPC} autoscaler, referred to as \emph{MPCScaler}. \ac{MPC} utilizes a system model to select actions that optimize future behavior under constraints and multiple objectives. In our setting, the system refers to the data center running the service, and the aim is to reduce energy consumption while maintaining latency within the \ac{SLO}.

At each control step $t$, the controller selects the replica count $R_t \in \mathbb{N}$ by solving an optimization over a finite horizon, using a surrogate model of performance and energy. Let $X_t$ be the observed or forecasted load at time $t$ in requests per second. For any candidate pair $(R, X)$, the surrogate predictor $\hat{g}$ returns the predicted P95 latency $\hat{L}(R,X)$ and energy consumption $\hat{E}(R,X)$:

\begin{equation}
\footnotesize
  \bigl( \hat{L}(R,X), \hat{E}(R,X) \bigr)
  = \hat{g}(R,X).
  \label{eq:surrogate}
\end{equation}

Given a forecast $\{X_{t+k}\}_{k=0}^{H-1}$ over a planning horizon $H = \texttt{cfg.horizon}$, the controller searches over sequences $\mathbf{R}_t = (R_t, R_{t+1}, \dots, R_{t+H-1})$ with $R_{\min} \le R_{t+k} \le R_{\max}$ for all $k$. For each candidate sequence, it minimizes the total cost
\begin{equation}
\footnotesize
  J(\mathbf{R}_t)
  =
  \sum_{k=0}^{H-1}
  \left(
    \alpha \, \hat{E}(R_{t+k}, X_{t+k})
    +
    \lambda_{\Delta} \, \bigl| R_{t+k} - R_{t+k-1} \bigr|
  \right),
  \label{eq:cost}
\end{equation}
where $\alpha = \texttt{cfg.energy\_gain}$ weighs the energy term,
$\lambda_{\Delta} = \texttt{cfg.lmbd\_drep}$ penalizes replica changes to promote stability, and $R_{t-1}$ is the replica count from the previous interval.

The plan must satisfy the latency \ac{SLO} at each step:
\begin{equation}
\footnotesize
  \hat{L}(R_{t+k}, X_{t+k})
  \le
  L_{\mathrm{SLO}}
  =
  \texttt{cfg.slo\_p95},
  \quad
  \forall k = 0, \dots, H-1,
  \label{eq:latency_constraint}
\end{equation}
and any sequence that violates \eqref{eq:latency_constraint} is infeasible.

Among feasible sequences, MPCScaler selects
$\mathbf{R}_t^\star = \arg\min J(\mathbf{R}_t)$ and applies only the first action $R_t^\star$.
At $t{+}1$, it updates measurements, refreshes predictions via \eqref{eq:surrogate}, and resolves the optimization in a receding horizon manner. The result is an adaptive autoscaling policy that is energy-aware, adheres to the \ac{SLO}, and balances energy use with scaling stability under the predicted workload.

\section{Sustainable data center Service Operation}\label{sec:uc1}


To showcase NEURONET and NeuroScaler, we target energy optimization in sustainable data center services. We deploy an \ac{HTTP}-based sensor service: clients send POST requests and receive \ac{HTTP} responses, which provide a real representation of the services we already operate across multiple data centers.

This service was deployed on a \ac{K8s} service cluster with three masters, three workers, and a load balancer, as depicted in Fig.~\ref{fig:neuronet-high-view} in a telco-graded data centre in ITAV, Portugal. This deployment is a typical \ac{K8s} cluster with \ac{HA}, which brings the experiment closer to a system in an operational environment. Moreover, by using \ac{K8s}, we can leverage energy consumption information from the \ac{PDU}, as well as from the bare metal servers using \ac{RAPL}~\cite{andringa_understanding_2025}, which Scaphandre exports~\cite{andringa_understanding_2025}. This enables us to gather information from both the server and the \acp{VM} that run on it. 

We can also grab energy consumption information from the containers by using \ac{Kepler}~\cite{andringa_understanding_2025}. This sensing capacity also includes sensors that obtain multiple types of computational data, such as \ac{CPU}, \ac{RAM}, disk, and network, provided by the server, \acp{VM}, and \ac{K8s} at the node and pod levels. This approach will also enable a correlation between energy consumption and several computing-related parameters from the virtualized infrastructure. 

To simulate the service's usage, we utilized the K6 benchmark tool, a \ac{K8s}-native tool for running, managing, and scaling performance tests using K6. K6 provided a stack that enables us to create the necessary jobs and pods for each test, manage the test lifecycle (including starting, monitoring, and cleanup), and log test results. This approach would also enable us to create conditions for test reproducibility using open-source software.

Figure~\ref{fig:neuronet-low-view} also presents, on the right side, the NEURONET \ac{K8s} cluster, including its components and highlights NeuroScaler. During experiment execution, NeuroScaler uses energy-consumption predictions to scale service replicas up and down.

\section{Experiments Description}
\label{sec:experiments-description}
This section describes the experiments in three different phases.


In the \textbf{first phase} (\emph{NO-AUTOSCALING}), we used k6 to apply a ramp load—gradually increasing and later decreasing \acp{VU}—against a single service replica. This profiled the \ac{K8s} test cluster’s compute and energy behavior and produced a dataset for \ac{ML} training via NEURONET’s \ac{MLOps} platform.

To make a representative dataset, we ran the first test for forty-eight (48) hours. The load pattern was a synodal load with a period of five (5) minutes, and the \acp{VU} variation was from zero (0) to one hundred (100). It is worth noting that the \ac{K8s} service cluster and the Neuronet cluster were deployed in a real data center at ITAV in Portugal. Thus, the data collection was conducted in a realistic environment where multiple workloads are running in parallel on different servers.


After the first phase of experiments, we execute the \textbf{\ac{ML}-model design} using NEURONET's \ac{MLOps} capabilities, implemented with open-source tools for design, implementation, and deployment of multiple \ac{ML} pipelines. These pipelines represent the core of the NEURONET \ac{AI} layer, enabling modular, reproducible, and fully automated workflows that cover the entire \ac{ML} lifecycle, from data retrieval and preprocessing to model training, hyperparameter optimization, and performance evaluation. Each pipeline is executed within a containerized \ac{K8s} environment, ensuring scalability, version control, and full traceability across experiment runs. All intermediate datasets, trained models, and evaluation metrics are automatically versioned and stored in MinIO, guaranteeing reproducibility and facilitating continuous experimentation. 

This setup enables continuous training and deployment of new or improved pipelines, supporting dynamic, energy-aware operations across the NEURONET infrastructure. The design details of these models and their comparison for improved fit to the use case are outside the scope of this work due to space constraints.

The following two phases are the experiments to compare the \ac{K8s} \ac{HPA} scalar with the MPC-based component of NEURONET's NeuroScaler (MPCscaler, presented in Subsection \ref{sec:mpc-autoscaler}).


In the \textbf{second phase} (\emph{K8S-AUTOSCALING}), we used the native K8S \ac{HPA}. It periodically reads built-in pod metrics (e.g., \ac{CPU}/memory), compares the average to a latency-derived target, and scales replicas up/down via proportional control, with rate/amount limits to prevent oscillations~\cite{nguyen_horizontal_2020}.

Finally, in the \textbf{third phase}, we used the \emph{NEURONET-AUTOSCALING}. In this phase, service scaling was based on the NeuroScaler autoscaling mechanism. This mechanism utilizes one of the previously designed, trained, and deployed \ac{ML} models. This model uses the features \emph{cpuload}, \emph{mem\_used}, and \emph{swap\_used} to predict the \emph{active power} at a given moment. Thus, by inspecting these service parameters, NeuroScaler can scale service replicas up or down, not only based on computing resources but also using a pre-trained \ac{ML} model focused on energy efficiency. This model, created by the initial data set obtained during the first phase, considers not only computing parameters from the \ac{K8s} tier but also energy and computing data from the different tiers presented in Fig~\ref{fig:neuronet-low-view}. By utilizing a richer computing dataset compared to \ac{K8s}, and incorporating energy consumption data from \ac{PDU}, Scaphandre, and \ac{Kepler} sensors, NEURONET advances to the state-of-the-art in energy efficiency-oriented autoscaling for \ac{K8s}-based environments \cite{beena_green_2025}.


The mechanism controls the number of replicas of a \ac{K8s} service. Unlike \ac{HPA}, which reacts to a single metric (e.g., \ac{CPU}), NEURONET's NeuroScaler applies \ac{MPC} to proactively choose the replica level that minimizes energy while meeting performance \acp{SLO}. We target P95 latency (the response time at or below which 95\% of requests complete), which reflects perceived speed for nearly all users. To run the tests, we use the control variables presented in Table \ref{tab:mpcscaler-params}.
\begin{table}[ht]
\centering
\footnotesize
\caption{MPCScaler configuration parameters used in the experiments.}
\label{tab:mpcscaler-params}
\setlength{\tabcolsep}{2pt}
\renewcommand{\arraystretch}{1.15}

\begin{tabularx}{\columnwidth}{
    >{\raggedright\arraybackslash}p{0.18\columnwidth}  
    >{\raggedright\arraybackslash}X                    
    >{\raggedright\arraybackslash}p{0.09\columnwidth}  
}
\toprule
\texttt{Variable} & \textbf{Description} & \textbf{Value} \\
\midrule
\texttt{COOL\_DOWN}
& Interval in seconds between scaling decisions.
& 15.0 \\
\texttt{SLO\_P95}
& P95 latency SLO in milliseconds.
& 1200.0 \\
\texttt{R\_MIN}, \texttt{R\_MAX}
& Minimum and maximum allowed replicas.
& [1, 20] \\
\texttt{HORIZON}
& Number of future steps in the MPC planning horizon.
& 3 \\
\texttt{LMBD\_DREP}
& Penalty weight for changing the replica count
; higher values enforce more stable scaling decisions.
& 4.0 \\
\bottomrule
\end{tabularx}
\end{table}

\section{Results and Discussion}\label{sec:results_and_discussion}

\begin{figure}[htpb]
\centering
  \includegraphics[width=0.90\columnwidth]{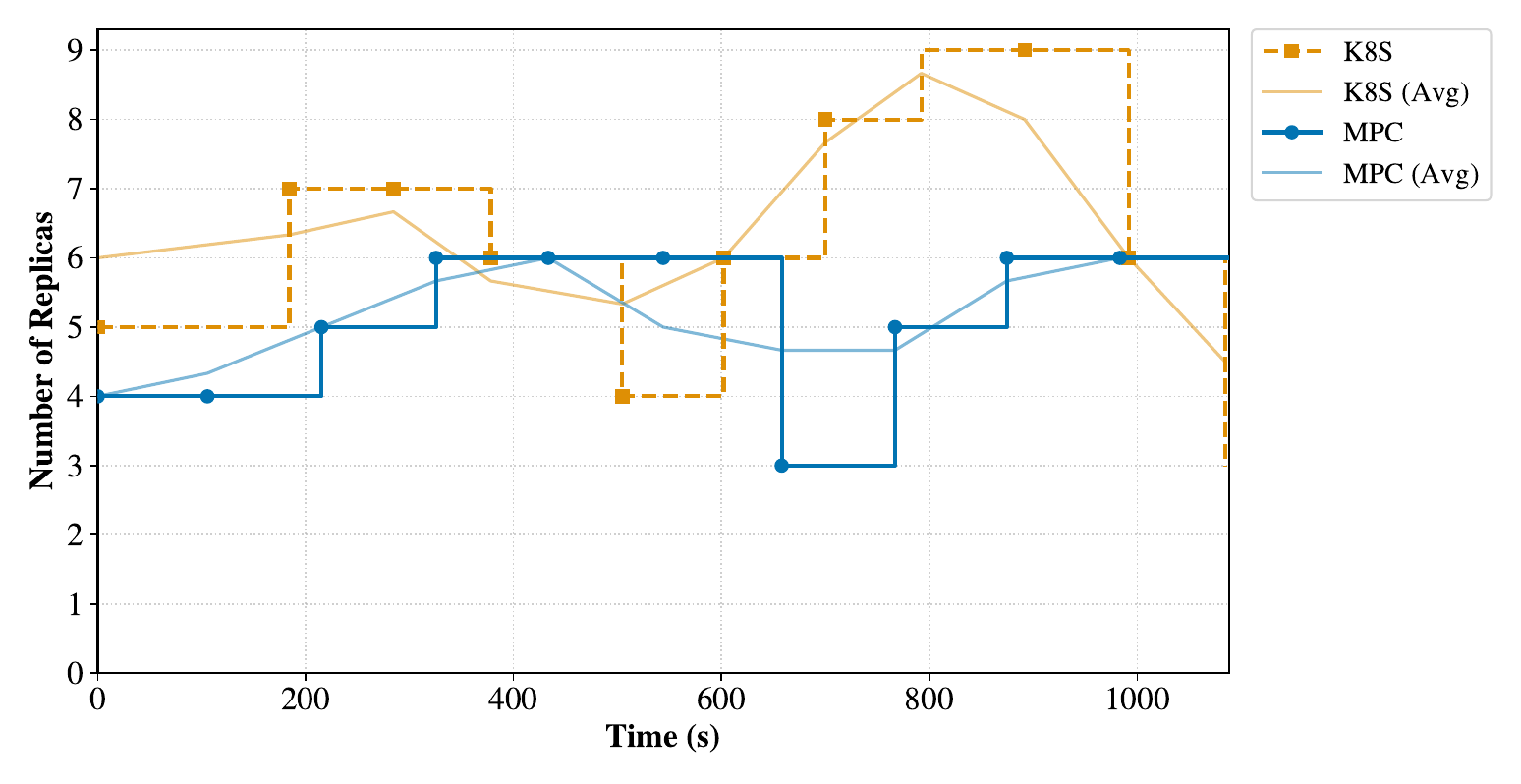}
  \caption{Replicas during experiments.}
  \label{fig:plot-replicas}
\end{figure}

\begin{figure}[htpb]
\centering
  \includegraphics[width=0.90\columnwidth]{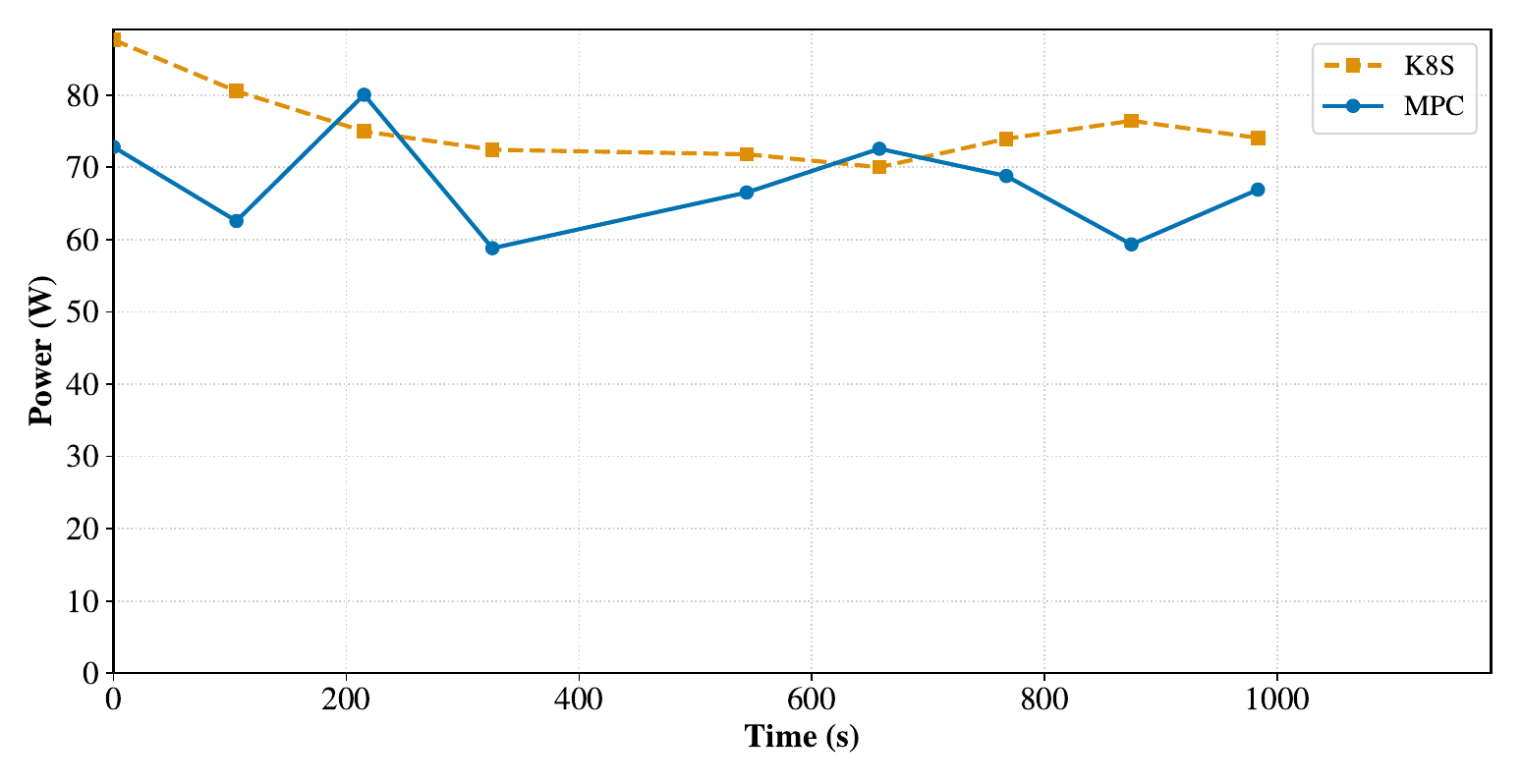}
  \caption{Power over time.}
  \label{fig:plot-energy}
\end{figure}

To test our approach, we subjected a web application to an identical, dynamic workload under two different autoscaling scenarios: our \ac{MPC}-based NEURONET autoscaler, used by NeuroScaler, and the standard \ac{K8s} \ac{HPA}. We measured key performance and energy metrics throughout using the same test window. Figure~\ref{fig:plot-replicas} presents the replicas in each scenario during the experiments, and Figure~\ref{fig:plot-energy} details energy consumption.

\begin{figure}[htpb]
\centering
\begin{subfigure}[b]{0.49\columnwidth}
  \centering
  \includegraphics[width=\linewidth]{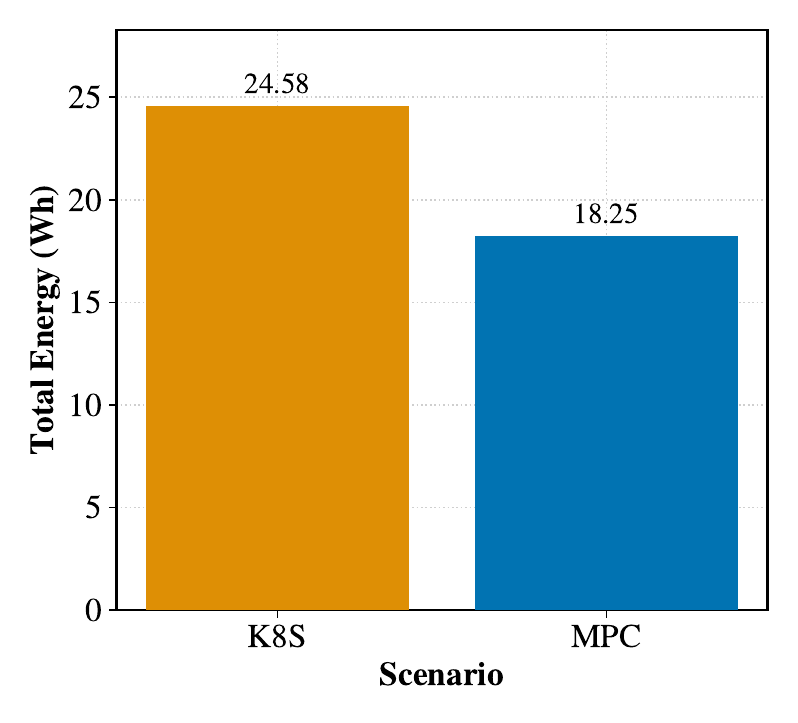}
  \caption{}
  \label{fig:plot-energy_bar}
\end{subfigure}
\hfill
\begin{subfigure}[b]{0.49\columnwidth}
  \centering
  \includegraphics[width=\linewidth]{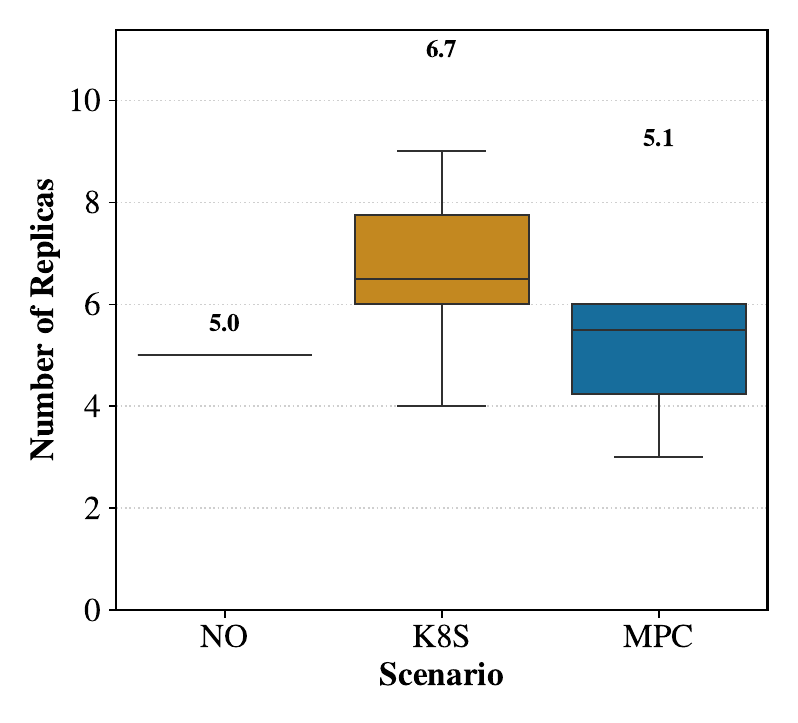}
  \caption{}
  \label{fig:plot-replica_ditribution}
\end{subfigure}
\caption{Resource usage comparison.}
\label{fig:energy_and_replica_distribution}
\end{figure}

The primary result is a substantial reduction in energy consumption. The \ac{MPC} controller's proactive and optimized scaling decisions lead to more efficient resource utilization, eliminating the waste associated with the \ac{HPA}'s tendency to over-provision replicas, as depicted in Figure~\ref{fig:plot-replica_ditribution}.

As depicted in the Fig.~\ref{fig:plot-energy_bar}, using the \ac{K8s} \ac{HPA} scaler, the service consumed 24.58 Wh to handle the workload imposed during the experimentation. Using the NEURONET Autoscaler, based on the \ac{MPC}, defined in the NeuroScalerMPC component, the active power consumption was 18.25 Wh for the same workload. In this case, the \ac{K8s} \ac{HPA} approach results in a 34.6\% increase in energy consumption compared to the value achieved with the NeuroScaler MPC-based approach.

As presented in Fig.~\ref{fig:plot-replica_ditribution}, using the \ac{K8s} \ac{HPA}, the average number of service replicas running was 6.7 during the experiments. On the other side, using the NeuroScaler, we observed an average of 5.1 replicas during the experiment. 

\begin{figure}[!ht]
\centering
\begin{subfigure}[b]{0.48\columnwidth}
  \centering
  \includegraphics[width=\linewidth]{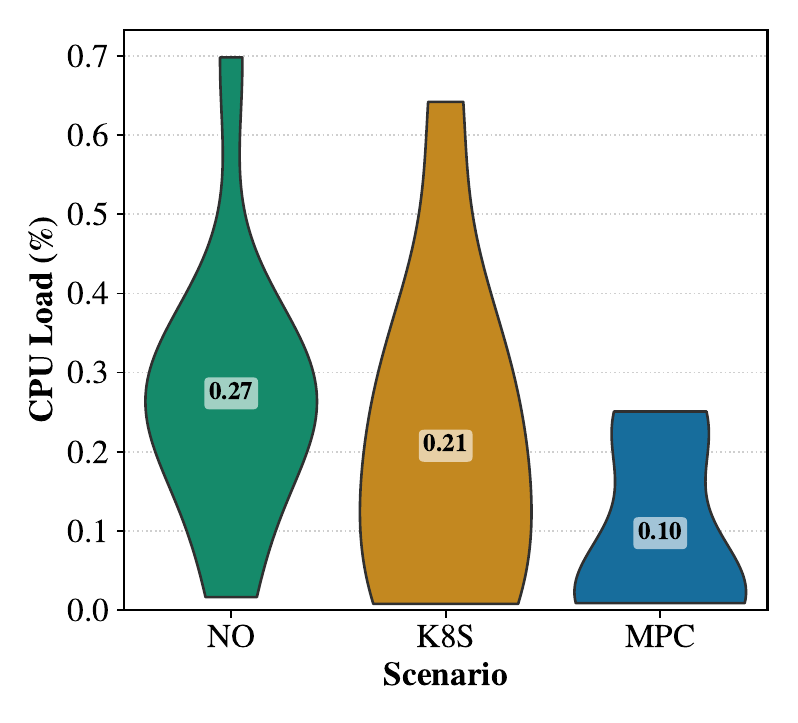}
  \caption{}
  \label{fig:plot-cpu_load}
\end{subfigure}
\hfill
\begin{subfigure}[b]{0.48\columnwidth}
  \centering
  \includegraphics[width=\linewidth]{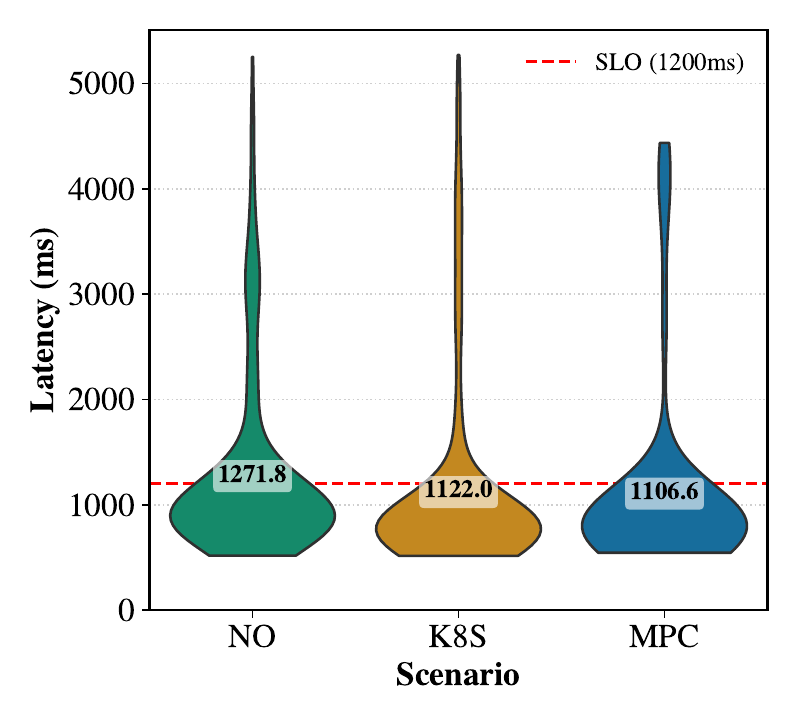}
  \caption{}
  \label{fig:plot-latency_load}
\end{subfigure}
\caption{Resource usage comparison.}
\label{fig:plot-resource-comparison}
\end{figure}

Figure \ref{fig:plot-cpu_load} shows that the average CPU load per pod using the \ac{K8s} \ac{HPA} scaler was 0.21, while the average CPU load per core using NeuroScaler was 0.10. The lower value indicates that, with NeuroScaler, each replica supported the workload with lower \ac{CPU} utilization. This indicator also helps to explain the energy efficiency in our approach. 

\section{Concluding Remarks}\label{sec:concluding_remarks}


This work introduces NeuroScaler, an AI-native, energy-aware autoscaling framework integrated into the NEURONET platform, to support greener telco clouds and data centres. Leveraging multi-layer telemetry and surrogate models, the proposed \ac{MPC}-based controller jointly optimizes resource allocation and energy consumption while preserving \acp{SLO} targets. 

Experiments on a telco-graded data center with realistic HTTP workloads in a production environment show that NeuroScaler can significantly reduce energy consumption compared with the standard \ac{K8s} \ac{HPA}, without violating latency \ ac {SLOs} or inducing instability. Using NeuroScaler, energy consumption is reduced by 34.68\% compared to Kubernetes native \ac{HPA} while maintaining target latency.

These results demonstrate that treating energy as a first-class control objective, supported by predictive and model-driven policies, is both practical and effective. 

Future work will present NeuroScaler for multi-service scenarios, with end-to-end optimization across data center, edge, and mobile network domains. Additionally, we plan to curate new \ac{ML} models that focus on other \acp{SLO} and usage scenarios.

\section*{Acknowledgment}


This work was supported by the EU Horizon Europe project IMAGINE-B5G (Grant No. 101096452) and FAPEMIG (Grant APQ-00923-24).

\bibliographystyle{IEEEtran}
\bibliography{references}

\end{document}

%% file: acronym.tex
\acrodef{3GPP}{3rd Generation Partnership Project}
\acrodef{5G}{5th Generation Mobile Network}
\acrodef{6G}{6th Generation Mobile Network}
\acrodef{AI}{Artificial Intelligence}
\acrodef{AI4Net}{\ac{AI} for Networking}
\acrodef{MDP}{Markov Decision Process}
\acrodef{AIDER}{Aerial Image Dataset for Emergency Response}
\acrodef{AMF}{Access and Mobility Management Function}
\acrodef{AIaaS}{Artificial Intelligence-as-a-Service}
\acrodef{AC}{Actor-Critic}
\acrodef{AIDER}{Aerial Image Database for Emergency Response applications}
\acrodef{API}{Application Programming Interface}
\acrodef{B5G}{Beyond Fifth Generation}
\acrodef{BPF}{Berkeley Packet Filter}
\acrodef{CBR}{Constant Bit Rate}
\acrodef{CSV}{Comma-Separated Values}
\acrodef{CPU}{Central Processing Unit}
\acrodef{CNN}{Convolutional Neural Network}
\acrodef{CNNs}{Convolutional Neural Networks}
\acrodef{C-V2X}{Cellular Vehicle to-Everything}
\acrodef{DoS}{Denial of Service}
\acrodef{DDQL}{Double Deep
 Q-learning}
\acrodef{DDoS}{Distributed Denial of Service}
\acrodef{DDPG}{Deep Deterministic Policy Gradient}
\acrodef{DNN}{Deep Neural Network}
\acrodef{DRL}{Deep Reinforcement Learning}
\acrodef{DQN}{Deep Q-Network}
\acrodef{DT}{Decision Tree}
\acrodef{DDQN}{Double Deep Q-Network}

\acrodef{ETSI}{European Telecommunications Standards Institute}
\acrodef{eNWDAF}{Evolved Network Data Analytics Function}
\acrodef{eBPF}{Extended Berkeley Packet Filter}
\acrodef{ECDF}{Empirical Cumulative Distribution Function}
\acrodef{ECDFs}{Empirical Cumulative Distribution Functions}
\acrodef{FIBRE}{Future Internet Brazilian Environment for Experimentation}
\acrodef{FL}{Federated Learning}
\acrodef{FEL}{Federated Ensemble Learning}
\acrodef{GNN}{Graph Neural Networks}
\acrodef{GPU}{Graphics Processing Unit}
\acrodef{GTP}{GPRS Tunnelling Protocol}
\acrodef{GTP-U}{GPRS Tunnelling Protocol User Plane}
\acrodef{GA}{Genetic Algorithm}
\acrodef{HTM}{Hierarchical Temporal Memory}
\acrodef{HPA}{Horizontal Pod Autoscaler}
\acrodef{HA}{High Availability}
\acrodef{HTTP}{Hypertext Transfer Protocol}
\acrodef{IAM}{Identity And Access Management}
\acrodef{ICMP}{Internet Control Message Protocol}
\acrodef{ICT}{Information and Communication Technologies}
\acrodef{IID}{Independent and Identically Distributed}
\acrodef{IoE}{Internet of Everything}
\acrodef{IoT}{Internet of Things}
\acrodef{ITU}{International Telecommunication Union}
\acrodef{IQR}{Interquartile Range}
\acrodef{I/O}{Input/Output}
\acrodef{IP}{Internet Protocol}
\acrodef{KNN}{K-Nearest Neighbors}
\acrodef{KPI}{Key Performance Indicator}
\acrodef{KPIs}{Key Performance Indicators}
\acrodef{K8s}{Kubernetes}
\acrodef{Kepler}{Kubernetes-based Efficient Power Level Exporter}
\acrodef{LSTM}{Long Short-Term Memory}
\acrodef{LOWESS}{Locally Weighted Scatterplot Smoothing}
\acrodef{LR}{Learning Rate}
\acrodef{MAE}{Mean Absolute Error}
\acrodef{MAD}{Median Absolute Deviation}
\acrodef{ML}{Machine Learning}
\acrodef{MLaaS}{Machine Learning as a Service}
\acrodef{MOS}{Mean Opinion Score}
\acrodef{MAPE}{Mean Absolute Percentage Error}
\acrodef{MSE}{Mean Squared Error}
\acrodef{MEC}{Multi-access Edge Computing}
\acrodef{mMTC}{Massive Machine Type Communications}
\acrodef{MFA}{Multi-factor Authentication}
\acrodef{MLP}{Multi-Layer Perceptron}
\acrodef{MADRL}{Multi-Agent Deep Reinforcement Learning}
\acrodef{MAB}{Multi-Armed Bandit}
\acrodef{MILP}{Mixed Integer Linear Programming}
\acrodef{MQTT}{Message Queuing Telemetry Transport}
\acrodef{MPC}{Model Predictive Control}
\acrodef{MLOps}{Machine Learning Operations}
\acrodef{NWDAF}{Network Data Analytics Function}
\acrodef{Net4AI}{Networking for \ac{AI}}
\acrodef{NS}{Network Slicing}
\acrodef{NFV}{Network Function Virtualization}
\acrodef{OSM}{Open Source MANO}
\acrodef{PCA}{Principal Component Analysis}
\acrodef{PC-FedAvg}{Personalized Conditional Federated Averaging}
\acrodef{PoC}{Proof of Concept}
\acrodef{PPO}{Proximal Policy Optimization}
\acrodef{POMDP}{Partially Observable Markov decision process}
\acrodef{PCAP}{Packet Capture}
\acrodef{PSO}{Particle Swarm Optimization}
\acrodef{PDU}{Power Distribution Unit}
\acrodef{PDUs}{Power Distribution Units}
\acrodef{QoE}{Quality of experience}
\acrodef{QoS}{Quality of Service}
\acrodef{QFI}{QoS Flow Identifier}
\acrodef{QFIs}{QoS Flow Identifiers}
\acrodef{RAM}{Random Access Memory}
\acrodef{RAN}{Radio Access Network}
\acrodef{RAPL}{Running Average Power Limit} 
\acrodef{RF}{Random Forest}
\acrodef{RL}{Reinforcement Learning}
\acrodef{RMSE}{Root Mean Square Error}
\acrodef{RNN}{Recurrent Neural Network}
\acrodef{RTT}{Round-Trip Time}
\acrodef{RAN}{Radio Access Network}
\acrodef{RTP}{Real-time Transport Protocol}
\acrodef{SDN}{Software-Defined Networking}
\acrodef{SFI2}{Slicing Future Internet Infrastructures}
\acrodef{SLA}{Service-Level Agreement}
\acrodef{SLO}{Service Level Objective}
\acrodef{SLOs}{Service Level Objectives}
\acrodef{SON}{Self-Organizing Network}
\acrodef{SMF}{Session Management Function}
\acrodef{S-NSSAI}{Single Network Slice Selection Assistance Information}
\acrodef{SVM}{Support Vector Machine}
\acrodef{SOPS}{Service-Aware Optimal
 Path Selection}
 \acrodef{SAFE}{Scalable Asynchronous Federated Ensembling}
\acrodef{TQFL}{Trust Deep Q-learning Federated Learning}
\acrodef{TEID}{Tunnel Endpoint Identifier}
\acrodef{TEIDs}{Tunnel Endpoint Identifiers}
\acrodef{TPE}{Tree-Structured Parzen Estimator}
\acrodef{UE}{User Equipment}
\acrodef{UEs}{User Equipments}
\acrodef{UPF}{User Plane Function}
\acrodef{UPFs}{User Plane Functions}
\acrodef{URLLC}{Ultra-Reliable and Low Latency Communications}
\acrodef{UAV}{Unmanned Aerial Vehicle}
\acrodef{UAVs}{Unmanned Aerial Vehicles}
\acrodef{UDP}{User Datagram Protocol}
\acrodef{VoD}{Video on Demand}
\acrodef{VR}{Virtual Reality}
\acrodef{AR}{Augmented Reality}
\acrodef{V2V}{Vehicle-to-Vechile}
\acrodef{V2X}{Vehicle-to-Everything}
\acrodef{VIM}{Virtual Infrastructure Manager}
\acrodef{VNF}{Virtual Network Function}
\acrodef{VNFs}{Virtual Network Functions}
\acrodef{VU}{Virtual User}
\acrodef{VM}{Virtual Machine}
\acrodef{XDP}{eXpress Data Path}

%% file: references.bib
@article{yrjola_sustainability_2020,
	title = {Sustainability as a Challenge and Driver for Novel Ecosystemic 6G Business Scenarios},
	volume = {12},
	rights = {http://creativecommons.org/licenses/by/3.0/},
	issn = {2071-1050},
	url = {https://www.mdpi.com/2071-1050/12/21/8951},
	doi = {10.3390/su12218951},
	pages = {8951},
	number = {21},
	journaltitle = {Sustainability},
	author = {Yrjölä, Seppo and Ahokangas, Petri and Matinmikko-Blue, Marja},
	urldate = {2024-12-07},
	date = {2020-01},
	langid = {english},
	note = {Number: 21
Publisher: Multidisciplinary Digital Publishing Institute},
}

@article{merluzzi_hexa-x_2023,
	title = {The Hexa-X Project Vision on Artificial Intelligence and Machine Learning-Driven Communication and Computation Co-Design for 6G},
	volume = {11},
	issn = {2169-3536},
	url = {https://ieeexplore.ieee.org/abstract/document/10156818},
	doi = {10.1109/ACCESS.2023.3287939},
	pages = {65620--65648},
	journaltitle = {{IEEE} Access},
	author = {Merluzzi, Mattia and Borsos, Tamás and Rajatheva, Nandana and Benczúr, András A. and Farhadi, Hamed and Yassine, Taha and Müeck, Markus Dominik and Barmpounakis, Sokratis and Strinati, Emilio Calvanese and Dampahalage, Dilin and Demestichas, Panagiotis and Ducange, Pietro and Filippou, Miltiadis C. and Baltar, Leonardo Gomes and Haraldson, Johan and Karaçay, Leyli and Korpi, Dani and Lamprousi, Vasiliki and Marcelloni, Francesco and Mohammadi, Jafar and Rajapaksha, Nuwanthika and Renda, Alessandro and Uusitalo, Mikko A.},
	urldate = {2024-12-07},
	date = {2023},
	note = {Conference Name: {IEEE} Access},
}

@article{nguyen_horizontal_2020,
	title = {Horizontal Pod Autoscaling in Kubernetes for Elastic Container Orchestration},
	volume = {20},
	rights = {http://creativecommons.org/licenses/by/3.0/},
	issn = {1424-8220},
	url = {https://www.mdpi.com/1424-8220/20/16/4621},
	doi = {10.3390/s20164621},
	pages = {4621},
	number = {16},
	journaltitle = {Sensors},
	author = {Nguyen, Thanh-Tung and Yeom, Yu-Jin and Kim, Taehong and Park, Dae-Heon and Kim, Sehan},
	urldate = {2025-10-18},
	date = {2020-01},
	langid = {english},
	note = {Publisher: Multidisciplinary Digital Publishing Institute},
}

@article{beena_green_2025,
	title = {Green Video Transcoding in Cloud Environments Using Kubernetes: A Framework With Dynamic Renewable Energy Allocation and Priority Scheduling},
	volume = {13},
	issn = {2169-3536},
	url = {https://ieeexplore.ieee.org/abstract/document/11071283},
	doi = {10.1109/ACCESS.2025.3585615},
	shorttitle = {Green Video Transcoding in Cloud Environments Using Kubernetes},
	pages = {116362--116390},
	journaltitle = {{IEEE} Access},
	author = {Beena, B. M. and Cheluvasai Ranga, Prashanth and Vinitha Chowdary, A. and Gamidi, Rohan and Hemasri, M. and Muppala, Tejaswi},
	urldate = {2025-10-18},
	date = {2025},
}

@inproceedings{caspian2024,
	title = {Caspian: A Carbon-aware Workload Scheduler in Multi-Cluster Kubernetes Environments},
	url = {https://ieeexplore.ieee.org/document/10786568},
	doi = {10.1109/MASCOTS64422.2024.10786568},
	shorttitle = {Caspian},
	eventtitle = {2024 32nd International Conference on Modeling, Analysis and Simulation of Computer and Telecommunication Systems ({MASCOTS})},
	pages = {1--8},
	booktitle = {2024 32nd International Conference on Modeling, Analysis and Simulation of Computer and Telecommunication Systems ({MASCOTS})},
	author = {Bahreini, Tayebeh and Tantawi, Asser N. and Tardieu, Olivier},
	urldate = {2025-11-10},
	date = {2024-10},
	note = {{ISSN}: 2375-0227},
}

@inproceedings{souza_casper_2024,
	location = {New York, {NY}, {USA}},
	title = {{CASPER}: Carbon-Aware Scheduling and Provisioning for Distributed Web Services},
	isbn = {979-8-4007-1669-0},
	url = {https://dl.acm.org/doi/10.1145/3634769.3634812},
	doi = {10.1145/3634769.3634812},
	series = {{IGSC} '23},
	shorttitle = {{CASPER}},
	pages = {67--73},
	booktitle = {Proceedings of the 14th International Green and Sustainable Computing Conference},
	publisher = {Association for Computing Machinery},
	author = {Souza, Abel and Jasoria, Shruti and Chakrabarty, Basundhara and Bridgwater, Alexander and Lundberg, Axel and Skogh, Filip and Ali-Eldin, Ahmed and Irwin, David and Shenoy, Prashant},
	urldate = {2025-11-10},
	date = {2024-05-29},
}

@article{carbonscaler2024,
	title = {{CarbonScaler}: Leveraging Cloud Workload Elasticity for Optimizing Carbon-Efficiency},
	volume = {7},
	url = {https://doi.org/10.1145/3626788},
	doi = {10.1145/3626788},
	shorttitle = {{CarbonScaler}},
	pages = {57:1--57:28},
	number = {3},
	journaltitle = {Proc. {ACM} Meas. Anal. Comput. Syst.},
	author = {Hanafy, Walid A. and Liang, Qianlin and Bashir, Noman and Irwin, David and Shenoy, Prashant},
	urldate = {2025-11-10},
	date = {2023-12-12},
}

@inproceedings{raith2024,
	title = {Opportunistic Energy-Aware Scheduling for Container Orchestration Platforms Using Graph Neural Networks},
	url = {https://ieeexplore.ieee.org/document/10701405},
	doi = {10.1109/CCGrid59990.2024.00042},
	eventtitle = {2024 {IEEE} 24th International Symposium on Cluster, Cloud and Internet Computing ({CCGrid})},
	pages = {299--306},
	booktitle = {2024 {IEEE} 24th International Symposium on Cluster, Cloud and Internet Computing ({CCGrid})},
	author = {Raith, Philipp and Rattihalli, Gourav and Dhakal, Aditya and Chalamalasetti, Sai Rahul and Milojicic, Dejan and Frachtenberg, Eitan and Nastic, Stefan and Dustdar, Schahram},
	urldate = {2025-11-10},
	date = {2024-05},
	note = {{ISSN}: 2993-2114},
}

@article{asadov2025,
	title = {Carbon-Aware Spatio-Temporal Workload Shifting in Edge–Cloud Environments: A Review and Novel Algorithm},
	volume = {17},
	rights = {http://creativecommons.org/licenses/by/3.0/},
	issn = {2071-1050},
	url = {https://www.mdpi.com/2071-1050/17/14/6433},
	doi = {10.3390/su17146433},
	shorttitle = {Carbon-Aware Spatio-Temporal Workload Shifting in Edge–Cloud Environments},
	pages = {6433},
	number = {14},
	journaltitle = {Sustainability},
	author = {Asadov, Nasir and Coroamă, Vlad C. and Franzil, Matteo and Galantino, Stefano and Finkbeiner, Matthias},
	urldate = {2025-11-10},
	date = {2025-01},
	langid = {english},
	note = {Publisher: Multidisciplinary Digital Publishing Institute},
}

@article{patel2024,
	title = {Modeling the Green Cloud Continuum: integrating energy considerations into Cloud–Edge models},
	volume = {27},
	issn = {1573-7543},
	url = {https://doi.org/10.1007/s10586-024-04383-w},
	doi = {10.1007/s10586-024-04383-w},
	shorttitle = {Modeling the Green Cloud Continuum},
	pages = {4095--4125},
	number = {4},
	journaltitle = {Cluster Computing},
	shortjournal = {Cluster Comput},
	author = {Patel, Yashwant Singh and Townend, Paul and Singh, Anil and Östberg, Per-Olov},
	urldate = {2025-11-10},
	date = {2024-07-01},
	langid = {english},
}

@InProceedings{moreira2023,
author="Moreira, Rodrigo
and Martins, Joberto S. B.
and Carvalho, Tereza C. M. B.
and Silva, Fl{\'a}vio de Oliveira",
editor="Barolli, Leonard",
title="On Enhancing Network Slicing Life-Cycle Through an AI-Native Orchestration Architecture",
booktitle="Advanced Information Networking and Applications",
year="2023",
publisher="Springer International Publishing",
address="Cham",
pages="124--136",
isbn="978-3-031-28451-9"
}

@ARTICLE{Martins2023,
  author={Martins, Joberto S. B. and Carvalho, Tereza C. and Moreira, Rodrigo and Both, Cristiano Bonato and Donatti, Adnei and Corrêa, João H. and Suruagy, José A. and Corrêa, Sand L. and Abelem, Antonio J. G. and Ribeiro, Moisés R. N. and Nogueira, José-marcos S. and Magalhães, Luiz C. S. and Wickboldt, Juliano and Ferreto, Tiago C. and Mello, Ricardo and Pasquini, Rafael and Schwarz, Marcos and Sampaio, Leobino N. and Macedo, Daniel F. and De Rezende, José F. and Cardoso, Kleber V. and De Oliveira Silva, Flávio},
  journal={IEEE Access}, 
  title={{Enhancing Network Slicing Architectures With Machine Learning, Security, Sustainability and Experimental Networks Integration}}, 
  year={2023},
  volume={11},
  number={},
  pages={69144-69163},
  doi={10.1109/ACCESS.2023.3292788}}

@ARTICLE{Pivoto2023,
  author={Pivoto, Diego Gabriel Soares and Rezende, Tibério Tavares and Facina, Michelle Soares Pereira and Moreira, Rodrigo and de Oliveira Silva, Flávio and Cardoso, Kleber Vieira and Correa, Sand Luz and Araujo, Antonia Vanessa Dias and Silva, Rogério Sousa E. and Neto, Heitor Scalco and de Lima Tejerina, Gustavo Rodrigues and Alberti, Antonio Marcos},
  journal={IEEE Access}, 
  title={{A Detailed Relevance Analysis of Enabling Technologies for 6G Architectures}}, 
  year={2023},
  volume={11},
  number={},
  pages={89644-89684},
  doi={10.1109/ACCESS.2023.3301811}}

@ARTICLE{Donatti2025,
  author={Donatti, Adnei Willian and Cristina Machado, Marcia and Alexander Lopez Martinez, Marvin and Rogério Antunes, Sabino S. and Carlos Figueiredo Souza, Eli and Corrêa, Sand L. and Ferreto, Tiago C. and Augusto Suruagy, José and Martins, Joberto S. B. and Cristina Carvalho, Tereza},
  journal={IEEE Access}, 
  title={{Energy Efficiency in Network Slicing: Survey and Taxonomy}}, 
  year={2025},
  volume={13},
  number={},
  pages={134570-134589},
  doi={10.1109/ACCESS.2025.3590365}}

@inproceedings{andringa_understanding_2025,
	location = {New York, {NY}, {USA}},
	title = {Understanding the Energy Consumption of Cloud-native Software Systems},
	isbn = {979-8-4007-1073-5},
	url = {https://dl.acm.org/doi/10.1145/3676151.3719371},
	doi = {10.1145/3676151.3719371},
	series = {{ICPE} '25},
	pages = {309--319},
	booktitle = {Proceedings of the 16th {ACM}/{SPEC} International Conference on Performance Engineering},
	publisher = {Association for Computing Machinery},
	author = {Andringa, Lars and Setz, Brian and Andrikopoulos, Vasilios},
	urldate = {2025-11-12},
	date = {2025-05-05},
}
